\documentstyle[12pt]{article}

\def\ol{\overline \Lambda}
\def\ms{ \overline {\rm MS}}
\def\lms{ \Lambda_{\ms}}
\def\rln{ {\rm ~ln}}

\def\ra{\rangle}
\def\la{\langle}

\begin{document}
\renewcommand{\Large}{\large}

\font\fortssbx=cmssbx10 scaled \magstep2
\hbox to \hsize{
\hskip.5in \raise.1in\hbox{\fortssbx University of Minnesota} 
\hfill\vtop{\hbox{\bf UMN-TH-1111/92}
            \hbox{(September 1992)}} }

\vspace{ .75in}

\begin{center}
{\large\bf Moments of the Virtual Photon Structure Function}
\\[.4in]
Sun Myong Kim and Thomas F. Walsh
\\[.4in]
{\it School of Physics and Astronomy, University of Minnesota \\
Minneapolis, Minnesota 55455}
\end{center}
%%%%%%%%%%%%%%%%%%%%%%%%%%%%%%%%%%%%%%%%%%%%%%%%%%%%%%%%%%%%%%%%%%%%%%%%
%...........................ABSTRACT...................................%
%%%%%%%%%%%%%%%%%%%%%%%%%%%%%%%%%%%%%%%%%%%%%%%%%%%%%%%%%%%%%%%%%%%%%%%%
\begin{abstract}
{
The photon structure function is a useful testing ground for QCD. It
is perturbatively computable apart from a contribution from
what is usually called the hadronic component of the photon.
There have been many proposals for this nonperturbative part of the 
real photon structure function. By studying moments of the virtual 
photon structure
function, we explore the extent to which these proposed nonperturbative
contributions can be identified experimentally. 
}
\end{abstract}

\setcounter{page}{1}
%%%%%%%%%%%%%%%%%%%%%%%%%%%%%%%%%%%%%%%%%%%%%%%%%%%%%%%%%%%%%%%%%%%%%%
%..........................INTRODUCTION..............................%
%%%%%%%%%%%%%%%%%%%%%%%%%%%%%%%%%%%%%%%%%%%%%%%%%%%%%%%%%%%%%%%%%%%%%%
\section{ Introduction }

With the discovery of asymptotic freedom, QCD has become the accepted
theory of hadrons.   However, at the energy scale of the formation of hadrons
(about a few GeV or less), perturbative QCD (PQCD) can  no longer be applied
alone, and nonperturbative QCD (NP) becomes important and may even rule.
In this regime, we have very limited tools we can use to understand
what is going on.

There is a fundamental dimensional constant giving a qualitative citerion of
separating these two energy regions, $\lms \equiv \ol$.  We use the modified
minimal subtraction renormalization scheme, $\ms$, throughout this paper.  We
can hope for some progress in understanding the interplay of nonperturbative and
perturbative QCD in those processes where both play a role. It may even be
important to  understand the relation of perturbative and nonperturbative
contributions in order to obtain $\ol$ from experiment. 

The photon structure function in 
$e^+ e^- \rightarrow e^+e^- +hadrons$ has been a useful tool to
study QCD. It has been calculated perturbatively to nonleading order,
which is necessary to obtain a meaningful comparison with QCD predictions
for other processes \cite{BB}. While valid asymptotically, the PQCD prediction
is modified at attainable $Q^2$ by the presence of a NP part, usually
called the ``hadronic" component of the photon. (This is an unfortunate
usage and we will refer in the following to the PQCD and NP contributions
to the photon structure function.) Our  point of view in this paper is that
the photon
structure function is valuable precisely because it contains both PQCD
and NP contributions at not too large $Q^2$. This may make it a unique
process in which we can study both of these as  functions
of the kinematic variables.

The most information can be extracted from this process by studying the
photon structure function for a virtual photon target whose invariant mass,
``$ - P^2$", can vary over a range from zero to some appreciable $P^2 \ll Q^2$. 

Here we exploit the moments of the photon structure function from the
calculation of Uematsu and Walsh for the perturbative part of the
struncture function \cite{UW}. (In particular, we use the normalization and
conventions of that paper.)  We use the suggestions of various authors for the
nonpurturbative part \cite{Newman}, \cite{NA3}, \cite{CD}.  This
``hadronic" part to the function has been obtained by these authors by fitting
experimental data from other processes. Typically, one finds the
pion structure function from a semihadronic process such as 
$\pi + p \rightarrow \mu^+ \mu^-  + anything$. The static quark model and
vector dominace are then used to obtain a conjectured NP component to
the photon structure function.

The basic idea of this paper is to examine the behavior of the total photon 
structure function (including both PQCD and NP parts) as a function of 
$Q^2$ and $P^2$, with the aim mentioned above of finding out how to 
distinguish the two experimentally. First we  describe models for the NP 
contribution--including the $Q^2$ dependence
often left out of consideration. We then discuss the real and virtual photon
structure functions. We then present our results for the $P^2$
behavior of the two components of the structure function. 

We are interested in the kinematic region
$2-4$ GeV$^2 \le Q^2 \le 50$ GeV$^2$ and $0 \le P^2 \le 1$ GeV$^2$. We would
like to stimulate experimental studies of this range of variables.

Throughout this paper, we set the quark mass to zero and the number of quark
flavors to four. While interesting and relevant to a comparison with experiment
at high $Q^2$, a consideration of quark mass effects here would distract from
our main point.

%%%%%%%%%%%%%%%%%%%%%%%%%%%%%%%%%%%%%%%%%%%%%%%%%%%%%%%%%%%%%%%%%%%%%%%%%%%
%................MODELS FOR THE NONPERTURBATIVE COMPONENT.................%
%%%%%%%%%%%%%%%%%%%%%%%%%%%%%%%%%%%%%%%%%%%%%%%%%%%%%%%%%%%%%%%%%%%%%%%%%%%

\section{Models for the Nonperturbative Component}

We think that it is very much an open question how the nonperturbative component
of the photon structure function is to be treated.  For the present, we describe 
the conventional view. We return to the question later in the paper. According
to this conventional view, the overall real photon structure function is a sum
of two parts,
\begin{equation}
F^{\gamma}_{2}(x,Q^2) = F^{\gamma}_{2, PQCD}(x,Q^2) + 
                        F^{\gamma}_{2,NP}(x,Q^2).
\label{xrtot}
\end{equation}
It is customary to write the nonperturbative contribution to the photon
structure function $F^{\gamma}_{2,NP}$ as a coherent sum of light vector 
meson states. From this vector meson dominance model we can then
identify the NP part of the photon structure function \cite{BW}, 
approximated by the $\rho$ meson contribution,
\begin{equation}
F^{\gamma}_{2,NP}(x,Q^2) \simeq \big( {\alpha \pi \over \gamma_\rho^2} \big)
    F_2^{\rho^0} (x,Q^2).
\label{F2h}
\end{equation}
Where $\gamma_\rho^2=f_\rho^2/\pi \simeq 2.2$ and $F_2^{\rho^0}$ can be
obtained from the pion structure function using the quark model \cite{BW},
\begin{equation}
F_2^{\rho^0}(x,Q^2) = F_2^{\pi^-} (x,Q^2) = x \big( {4 \over 9} f_{{\bar u}
  /\pi^-} + {1 \over 9} f_{ d / \pi^-} \big) = {5 \over 9} x f_{{\bar u}
  /\pi^-}.
\end{equation}
Here we denote $x f_{ {\bar u}/\pi^-}$ as $x \bar u(x,Q^2)$ for
convenience and similarly for the others.
We also put ${\bar u}(x,Q^2)=d(x,Q^2)$ in the pion.  This assumes that we are at
low enough $Q^2$ so that we can neglect sea quarks. Then equation (\ref{F2h})
becomes
\begin{equation}
F^{\gamma}_{2,NP}(x,Q^2) = \big( {\alpha \pi \over \gamma_\rho^2} \big)
  {5 \over 9} x \bar u(x,Q^2).
\end{equation}
From now on we confine ourselves to $Q^2 \le 50$ GeV$^2$ where
we expect that our neglect of sea quarks will be adequate.
From definitions of the non-singlet and  the singlet quark distribution
functions we can write a quark structure function in terms of the non-singlet
and singlet parts of the structure function \cite{BP}.
\begin{equation}
x q_i (x,Q^2) = x q^{NS}_i (x,Q^2) + {1 \over 2 f} x q^S (x,Q^2),
\end{equation}
where $q^S \equiv \sum_i^f ( q_i + \bar q_i )$ and $i$ runs over all available
flavors, here we choose the number of flavors to four, $f=4$.
The final form of $F^{\gamma}_{2,NP}$ of the photon structure
function becomes
\begin{equation}
F^{\gamma}_{2,NP} = \big( {\alpha \pi \over \gamma_\rho^2} \big)
  {5 \over 9} \big[ x q_i^{NS}(x,Q^2) + {1 \over 2f} x q^S(x,Q^2) \big].
\end{equation}
We have $x q^{NS}_i \equiv x \bar u^{NS}$ for the pion in this example. The
moments of the nonperturbative part of the structure function are then
\begin{equation}
F_{2,NP}^\gamma (n, Q^2) = \int^1_0 dx x^{n-2}  F^{\gamma}_{2,NP} (x,Q^2).
\end{equation}
Our procedure will be to obtain $x q^{NS}$ and $x q^S$ at all $Q^2$ through
numerical integration of the Altarelli-Parisi equations.
So we can calculate $ M^\gamma_{n,NP} (Q^2)$ at any $Q^2$, given its
starting value. We now need to determine the starting value of the $\rho$
meson or pion structure function at some low $Q_0^2$.

We use three different fits
which were obtained by three groups Newman et. al. \cite{Newman}, NA3
collaboration at CERN \cite{NA3}, and Castorina et al. \cite{CD}.   All these
fits have been obtained at $Q^2 \simeq 2 - 4$ GeV$^2$. We will show all the 
solutions to the Altarelli-Parisi equations for the
corresponding NP contribtutions to the structure function. 
We emphasize that we have no convictions
as to which of the resulting NP parts is the correct one. Our aim later  in this
paper will be to see to what extent they can be distinguished experimentally for
$P^2 > 0$.

We will include three examples
of Non-Singlet (NS), Singlet (S), Gluon (G) parts of the structure function
at $Q^2_0$.

The general form of the parametrizations of the hadronic functions can be
written as $ x \bar u = C x^a (1-x)^b $ since the three fits are given as
\begin{equation}
\cases{ 
\rm (A) ~Newman ~et ~al. &: $x \bar u(x,Q_0^2)= 0.52 (1-x)$ \cr
\rm (B) ~NA3 ~collabaration &: $x \bar u(x,Q_0^2)= 0.57 x^{0.4} (1-x)$ \cr 
\rm (C) ~Castorina ~et ~al. &: $x \bar u(x,Q_0^2)= 0.766 x^{0.5}(1-x)^{0.6}$\cr
}
\end{equation}
which yield the nonperturbative structure functions considering only the
dominant $\rho$ meson
\begin{equation}
F^{\gamma}_{2,NP}(x,Q^2_0) =
\cases{ 
0.13~\alpha~(1-x)               &(A) \cr
0.14~\alpha~x^{0.4} (1-x)       &(B) \cr 
0.19~\alpha~x^{0.5} (1-x)^{0.6} &(C).\cr
}
\end{equation}

The first of these is slightly at odds with our neglect of sea quarks,
since it goes to a finite limit as $x \rightarrow 0$.
Since our aim is only to compare different assumptions about the $\rho$ meson
structure function for high moments, we will use it anyway.

Let us consider the ratios of the momenta carried by quarks and gluons.
Since $\bar u = d$ inside the $\pi^-$ at low $Q^2$, we can only consider the
$\bar u$ for quarks.  We define the ratios for the quarks and gluons as
\begin{eqnarray}
r_{\bar u} =& \int_0^1 dx x f_{\bar u/\pi^-} =& \int_0^1 dx x \bar u(x,Q^2)
\nonumber\\
r_g =& \int_0^1 dx x f_{g/ \pi^-} =& \int_0^1 dx x g(x,Q^2).
\end{eqnarray}
We can calculate $r_{\bar u}$ using the three fits
\begin{equation}
r_{\bar u} = C \int_0^1 dx x^a (1-x)^b = B(a+1, b+1),
\end{equation}
where $B$ is the Euler Beta function with $Re~(a+1) > 0$, $Re~(b+1) > 0$.
Therefore the values of $r_{\bar u}$ for each fit are
\begin{equation}
r_{\bar u} = \cases
{ 0.52 ~B(1,2)      &= 0.26  ~~(A)\cr
  0.57 ~B(1.4,2)    &= 0.17  ~~(B)\cr
  0.766 ~B(1.5,1.6) &= 0.28  ~~(C).\cr
}
\end{equation}
From the QCD momentum sum rule we have $r_g=1 -2r_{\bar u}$ so that 
\begin{equation}
r_g=1 - 2~C~B(a+1,b+1)= \cases
{ 0.48 \quad(A) \cr
  0.66 \quad(B) \cr
  0.44 \quad(C).\cr
}
\end{equation}
We can determine $r_g$ directly from $r_{\bar u}$ with the assumption of no
sea quark contribution to the total momentum at the low initial value of $Q^2$
regardless of the form of the gluon structure function.  The parametrization of
$xg(x,Q^2_0)$ is up to us to determine.  The simplest form of the gluon structure
function at $Q^2_0$ is
\begin{equation}
xg(x,Q_0^2) = C_1 (1-x)^{C_2}.
\label{GluSt}
\end{equation}
We can find the value of $C_1$ from the QCD momentum sum rule after making an
assumption of the value of $C_2$ ($C_2 \geq 0$) which could in principle
be determined from experiment.  We choose $C_2=2$ here.
Using the momentum sum rule again we obtain
\begin{equation}
r_g={C_1 \over (1+C_2)} = 1 - 2~C~B(a+1,b+1).
\end{equation}
This equation yields
\begin{equation}
C_2=2, \quad C_1=
\cases{ 0.624 &(A) \cr
        1.584 &(B) \cr
        1.344 &(C).\cr }
\end{equation}
%
%Here we have a freedom to choose $C_2$ and then determine $C_1$.

Let us look at the values of $r_{\bar u}$ for the three examples.
In example (A) 26\% of the total momentum is carried by the $\bar u$ quark and
thus 48\% of it is carried by gluons in $\rho$ (or $\pi$)-meson. 
Similar agument applies to other examples although, with the model that we
suggest, the example (B) seems to have too big gluon momentum (66\%) compared to
quarks (34\%).  The authors of reference (C) chose
$xg(x,Q_0^2)=1.408(1-x)^3$.  Then $r_g=0.35$ which seems small compared to our
result $0.44$. However, one needs to keep in mind that the model for the gluon
structure function here is only the simplest one and all three fits to the
experimental data are perfectly acceptable.

This analysis is independent of the number of vector mesons.  However, in
the nonperturbative part of the structure function we need to sum over all the
vector mesons coherently.  For example, we will have about 60\% enhancement in
the nonperturbative part when we include the incoherent sum of all vector mesons
$\omega$ and $\phi$ in addition to the dominant $\rho$ \cite{BW}. 
In this paper we confine ourselves only to the $\rho$-term for simplicity.

In the numerical calculation, we evolve the equations from $Q_0^2=3$ GeV$^2$ to
$Q^2=45$ GeV$^2$ with $\ol=0.2$ GeV.  The evolution of the nonsinglet function
can be obtained in a straight forward manner using the above hadronic functions.
In coupled equations of singlet and gluonic functions
we assume the initial gluon distribtuion in the form of
equation (\ref{GluSt}) with no initial sea quark distribution. We use the
normalization factor $\alpha \rln (Q^2/\ol^2)$ and we cut at very low $x$ ($x <
0.025$) and  high $x$ ($x > 0.975$) to avoid numerical
problems at the end points.

We show the moments of the NP part of the structure function,
$F_{2,NP}^\gamma (n,Q^2)$, in Figure~\ref{NPmoments}
with the normalization factor, $\alpha \rln (Q^2/\ol^2)$.
The qualitative behavior of NP of three fits are
same.  They decrease fast at low $Q^2$ while the rate of the decrement becomes
small at high $Q^2$.  This means that we do not expect very small NP
contribution at a given moment to the structure function even at high $Q^2$. 
Therefore we expect the NP contribution to the total structure function
persists even at high $Q^2$ as we will see in the next section.  We will come
back to this point in the real and the virtual cases.

%%%%%%%%%%%%%%%%%%%%%%%%%%%%%%%%%%%%%%%%%%%%%%%%%%%%%%%%%%%%%%%%%%%%%%%%%%%%%
%...................THE REAL PHOTON STRUCTURE FUNCTION......................%
%%%%%%%%%%%%%%%%%%%%%%%%%%%%%%%%%%%%%%%%%%%%%%%%%%%%%%%%%%%%%%%%%%%%%%%%%%%%%

\section{ The Real Photon Structure Function  }

The PQCD moments of the real photon structure function can be written as
\begin{eqnarray}
F_{2, PQCD}^\gamma (n,Q^2) =
\int_0^1 dx x^{n-2} F_2^\gamma (x,Q^2)
 &=& {\alpha^2 \over e^2} \big( {16 \pi^2 \over \beta_0 \overline g ^2} a_n
     + b_n \big) \cr
 &\simeq& {\alpha^2 \over e^2} \big( a_n \rln {Q^2 \over
   {\overline{\Lambda}}^2 } + \tilde a_n \rln \rln {Q^2 \over
   {\overline{\Lambda}}^2} + b_n \big).
\label{rPSF}
\end{eqnarray}
Where the expressions and the values of $a_n$, $b_n$, and $\tilde a_n$ are given
in reference \cite{BB}. The factor $1/e^2$ in the right hand side of the
equation is from the difference in normalizations of structure functions between
reference \cite{UW} on which our calculation is
based  and reference \cite{BB}. The approximation comes from omitting the higher orders in
\begin{equation}
\beta_0 \rln {Q^2 \over \ol^2} ~O \bigg[ \bigg( { \beta_1 \over
\beta_0^2}  {\rln \rln (Q^2/\ol^2) \over \rln (Q^2/\ol^2)} \bigg)^2
\bigg].
\end{equation}
$\ol$ is given in the $\ms$ scheme and
$\overline g ^2(Q^2) \equiv \overline g ^2_{\ms} (Q^2)$ is the
coupling constant with two-loop order term,
\begin{equation}
{\overline g ^2 (Q^2) \over 16 \pi^2}
= {1 \over { \beta_0 \rln (Q^2 / \ol^2) } }
   - { {\beta_1 \rln \rln (Q^2 / \ol^2) } \over 
        { \beta_0^3 \rln^2 (Q^2 / \ol^2) } }.
\label{gs}
\end{equation}
Let us go back to the equation (\ref{rPSF}); then we have
\begin{equation}
\tilde a_n = {\beta_1 \over \beta_0^2} a_n.
\end{equation}

We split the total
structure function into two parts as before. 
We use the results in \cite{UW} for the
perturbative QCD part of the structure funciton and the previous three examples
for the NP part.  We can recover Bardeen and Buras's moments from Uematsu and
Walsh's ones formally by putting $P^2=\ol^2$.

We write the moments of the real photon structure function
in equation (\ref{xrtot}) as 
\begin{equation}
F_2^\gamma (n,Q^2)=F_{2,PQCD}^\gamma (n,Q^2) + F_{2,NP}^\gamma (n,Q^2).
\label{nrtot}
\end{equation}

We obtain the total real photon structure
function  in each moment by summing the nonperturbative and
the perturbative contribution for the corresponding expressions.
As we saw in Figure \ref{NPmoments}, the NP parts of the structure functions
decrease relatively fast at low $Q^2$ while the rate slows down as $Q^2$
increases.   On the other hand PQCD increases fast at low $Q^2$ and slowly at
high $Q^2$ as we see in Figure \ref{realcase}.
This is a reasonable result since we expect
that NP contribution becomes larger than that of PQCD at a low $Q^2$.
If we add the two contributions, the total structure functions,
$F_{2,tot}^{\gamma , real} (n,Q^2)$,
for $n=4$ increase very slowly at high $Q^2$ after
normalizing by the factor $\alpha \rln (Q^2/ \ol^2)$.
Here the dashed curve indicates the PQCD part and A, B, and C indicate the total
structure functions for the three examples.
All the moments (for $n > 4$) increase rather slowly after $Q^2 > 15$
GeV$^2$.  We choose different scales for Figure \ref{realcase}(a) and
Figure \ref{realcase}(b) to distinguish explicitly the curves for those
examples.
 We also notice the faster decrease of the NP part compared to PQCD at higher
moments. The total real structure functions for all three fits are more or less
the same. We found that the NP contribution to
the total structure function is about 10\% around $Q^2=3$ GeV$^2$ when $n=4$ and
slightly less when the moments become higher and $Q^2$ gets larger.

%%%%%%%%%%%%%%%%%%%%%%%%%%%%%%%%%%%%%%%%%%%%%%%%%%%%%%%%%%%%%%%%%%%%%%%%%%%%%
%.......................VIRTUAL PHOTON STRUCTURE FUNCTION...................%
%%%%%%%%%%%%%%%%%%%%%%%%%%%%%%%%%%%%%%%%%%%%%%%%%%%%%%%%%%%%%%%%%%%%%%%%%%%%%

\section{ Virtual Photon Structure Function  }

Let us now move on to a virtual target photon. Consider a deep inelastic 
scattering involving two photons with  momenta
$q$ for the probe photon and $p$ for the target photon.   
We have $q^2=-Q^2$ and  $p^2=-P^2$ as the probe photon mass and the target
photon mass respectively. We take the kinematic region of the momenta of the
photons as follows.
\begin{equation}
\ol^2 \ll P^2 \ll Q^2.
\label{KinCon}
\end{equation}
The moments of the structure function can be calculated
by solving the renormalization group equation for the Wilson coefficient
in the operator product expansion of photon-photon scattering to leading order
in $\alpha$ and to next to leading order in $\alpha_s$.
\begin{eqnarray}
F_{2, PQCD}^\gamma (n,Q^2,P^2)
 &\equiv& \int_0^1 dx x^{n-2} F_2^\gamma (x,Q^2,P^2)  \cr
 &=&  {1 \over {16 \pi^2}} {e^2 \over {2 \beta_0}} 
   \bigg[ \sum_i \tilde P_i^n {1\over {1+\lambda_i^n/2 \beta_0} } 
   {{16\pi^2} \over {\overline g ^2(Q^2)} } 
   \bigg\{ 1- \bigg( 
   { {\overline g ^2(Q^2)} \over {\overline g ^2(P^2)} } \bigg)^ 
   { \lambda_i^n/2 \beta_0 +1 } \bigg\} \cr
 && \qquad\qquad + \sum_i A_i^n \bigg\{ 1- \bigg( 
   { {\overline g ^2(Q^2)} \over {\overline g ^2(P^2)} } \bigg) ^ 
       { \lambda_i^n/2 \beta_0 } \bigg\}   \cr
 && \qquad\qquad + \sum_i B_i^n \bigg\{ 1- \bigg( 
   { {\overline g ^2(Q^2)} \over {\overline g ^2(P^2)} } \bigg) ^ 
       { \lambda_i^n/2 \beta_0 +1 } \bigg\}  
       + C_\gamma^n \bigg]  ,
\label{momNLO}
\end{eqnarray}
where $\tilde P_i^n,~A_i^n,~B_i^n$, $~C_\gamma^n$, and $\lambda^n_i$ are given
in references \cite{UW}, \cite{err} and the index $i$ runs over $+$, $-$, NS.
Notations, $+$, $-$, NS are from the eigenvalues, $\lambda^n_{\pm}$,
$\lambda^n_{NS}$, of the one-loop hadronic anomalous dimension matrix.
$~~\overline g ^2(Q^2)$ is the coupling constant to two loops.   Now, equation
(\ref{momNLO}) is valid for $n=2$ since the singular behavior of the
coefficient $A^n_-$, which causes a problem in the real photon structure
function, is cancelled by the factor,
$1-(\bar g^2(Q^2)/ \bar g^2(P^2))^{\lambda^n_-/2 \beta_0}$
when $\lambda^n_-$ goes to zero.
Since the expression (\ref{momNLO}) is independent of any renormalization
scheme, to change the equation from one scheme to another all we have to do is
to write $\overline g$, $A_i^n$,
$B_i^n$, and $\Lambda$ in the new scheme.

We now write
\begin{equation}
F_2^\gamma (x,Q^2,P^2)
= F_{2,PQCD}^\gamma (x,Q^2,P^2) + F_{2,NP}^\gamma (x,Q^2,P^2)
\label{xvtot}
\end{equation}
with the corresponding moment functions
\begin{equation}
F_2^\gamma (n,Q^2,P^2)
= F_{2,PQCD}^\gamma (n,Q^2,P^2) + F_{2,NP}^\gamma (n,Q^2,P^2).
\label{nvtot}
\end{equation}
We continue to follow the convention and the simple vector meson dominance
model for the nonperturbative part of the structure function by writing
its dependence on $P^2$ as
\begin{equation}
F_{2,NP}^\gamma (x,Q^2,P^2)
={ F_{2,NP}^\gamma (x,Q^2) \over (1+ {P^2 \over M_\rho^2})^2 }.
\label{Fnp}
\end{equation}
Where $F_{2,NP}^\gamma$ is the real photon nonperturbative part which we have
been discussing up to this point. In the previous section we simply had $P^2=0$.
This is the only consistent way of including the $P^2$ dependence once
we adopt the notion that the NP part of the photon structure function can
be approximated by the simple vector dominance model.

As $n$ increases, the moments are sharply
suppressed in exactly the same way as for the real photon case.
Note that the
$n=2$ case is possible for both the nonperturbative and perturbative parts
in the virtual photon structure function while only the nonperturbative part
appears in the real structure function case.

Since the PQCD structure function of the real photon can be obtained from the virtual
photon structure function formally by putting $P^2=\ol^2$, it makes sense to
replace $P^2$ in our expressions by $P^2 + \ol^2$. Then we return to the real
photon structure function by setting $P^2=0$ rather than $\ol^2$. 
This modification does not affect our
calculation much since  we need  $P^2=0$ and
$P^{2} \gg \ol^2$. Henceforth we write the PQCD piece with this
change.

In Figure \ref{total} we show the total virtual photon structure function,
$F_{2, tot}^{\gamma, vir} (n,Q^2,P^2)$,
versus the probe photon mass $Q^2$ from PQCD and NP after normalizing the
structure function by $\alpha \rln (Q^2/({P}^2+\ol^2))$.
These display
the strong nonperturbative dependence of the total structure function at low
$Q^2$ ($Q^2 < 10$ GeV$^2$) and at low moments.   Perhaps the most interesting
behavior of the function occurs in the
regions $0.4$ GeV$^2 \le P^2 \le 0.6$ GeV$^2$ and $Q \ge 20$ GeV$^2$.
As the moments increase we observe similar behavior to that in the real
photon case.

It is useful to look at the $n=2$ moment in the virtual structure function since
there is no such moment in the real case. In fact we found interesting behavior
of the structure functions in our three examples.
When $P^2=0.2$ GeV$^2$, we have qualitatively same behavior for all three fits
while the fit (A) behaves very differently when $P^2=0.4$ GeV$^2$.
However, one should keep in mind that we have used only $\rho$ meson in the
analysis.  

From these plots we can clearly see how the NP contribution behaves, given
different assumptions about its form at $P^2=0$.

We close with a comment on the PQCD transition to the real photon case.
The virtual photon structure function can be reduced to the real structure
function by taking the target photon mass equal to zero
(i.e., $P^2=\ol^2$ in equation (\ref{momNLO})). 
Then $ \overline g^2(P^2=\ol^2) $ goes to the infinity where quark confinement
start occurring.  The equation (\ref{momNLO}) becomes
\begin{eqnarray}
&&\int_0^1 dx x^{n-2} F_2^\gamma (x,Q^2,P^2=\ol^2) \cr
         &&= {1 \over {16 \pi^2}} {e^2 \over {2 \beta_0}} 
            \bigg[~ \sum_i \tilde P_i^n {1\over {1+\lambda_i^n/2 \beta_0} } 
              {{16\pi^2} \over {\overline g ^2(Q^2)} }
            + \sum_i A_i^n + \sum_i B_i^n  
            + C_\gamma^n \bigg].
\label{rvPSF}
\end{eqnarray}
Now this expression is valid only for $n > 2$ since we cannot calculate the
moment for $n=2$ case in the real structure function due to the existence of
unknown hadronic matrix elements.  If we compare two equations (\ref{rPSF})
and (\ref{rvPSF}) in the $\ms$ scheme ($\ol$ and $\overline g ^2_{\ms}$ are
defined in the equation (\ref{gs})),
\begin{equation}
 {\alpha^2 \over e^2} \big( {16 \pi^2 \over \beta_0 \overline g ^2} a_n
     + b_n \big)
 ={1 \over {16 \pi^2}} {e^2 \over {2 \beta_0}} 
  \bigg[~ \sum_i \tilde P_i^n {1\over {1+\lambda_i^n/2 \beta_0} } 
   {{16\pi^2} \over {\overline g ^2(Q^2)} }
    + \sum_i A_i^n + \sum_i B_i^n  
   + C_\gamma^n \bigg].
\end{equation}
From the above equation we can easily identify
\begin{eqnarray}
  a_n &=& {1 \over 2\beta_0}~ \sum_i \tilde P_i^n {1\over {1+\lambda_i^n / 2
    \beta_0} } {{16 \pi^2} \over {\overline g ^2(Q^2)} } \cr  
  b_n &=& {1 \over {2 \beta_0}} (\sum A_i^n +\sum B_i^n+ C_\gamma^n ).
\end{eqnarray}
Therefore we obtain all coefficients in Bardeen and Buras' form from
those in Uematsu and Walsh.

%%%%%%%%%%%%%%%%%%%%%%%%%%%%%%%%%%%%%%%%%%%%%%%%%%%%%%%%%%%%%%%%%%%%%%%%
%..........................DISCUSSION..................................%
%%%%%%%%%%%%%%%%%%%%%%%%%%%%%%%%%%%%%%%%%%%%%%%%%%%%%%%%%%%%%%%%%%%%%%%%

\section{ Discussion }

The nonperturbative part of the photon structure function has been of interest
for a long time. At low $Q^2$ it complicates efforts to check QCD experimentally
and to extract the scale factor $\ol$. These problems can be dealt with in part
by studying the photon structure function for higher moments where the
prediction of QCD is stable against the higher order corrections \cite{BB} and
where the nonperturbative part is small. (Of course this depends at least in
part on the common assumption that the NP part vanishes quickly as the scaling
variable $x \rightarrow 1$.) The agreement of PQCD with experiment and a value
$\ol \simeq 200$ MeV is striking \cite{BW}. Nevertheless, it is still of
interest to know  to what extent the NP part of the structure function is
really under control at low $Q^2$. 

We have argued here that one should explore the dependence of the virtual photon 
structure function on target $P^2$. Then one can determine to an adequate extent
just how important the NP part is. We think that this is clearly possible,
given our results for the several NP models which we have studied.

This has a number of consequences. First, by
learning the actual size of $F_{NP}$, we can improve checks on the PQCD
prediction  and extract a more reliable value for $\ol$. Secondly, the NP part
and its  relationship to the conventional PQCD prediction is of interest by
itself. We know that checks of PQCD for high moments are satisfactory. However,
we also know that we have no prediction for the $n=2$ moment at $P^2 = 0$. It
is a surprising and interesting fact that the virtual photon structure function
has a calculable $n=2$ moment. If we want to understand the relationship of
perturbative and nonperturbative QCD, this problem needs to be explored
experimentally. Equivalently, we need to understand the behavior of the
structure function as $x \rightarrow 0$ for both real and virtual photon
targets.  The existence of a finite or logarithmically rising
cross section for the center of mass energy $W \rightarrow \infty$  at fixed
$P^2 \ll Q^2$ is a presumably nonperturbative phenomenon. This is related in
some yet unknown way to the relation of perturbative and nonperturbative
contributions to the structure function for small x. We think that this is an
important issue. It is as yet unexplored by experiment.

We can illustrate one unresolved issue by returning to our bland
assumption that the NP part of the structure function can be obtained from
the vector meson dominance model. While we have used this assumption to show
that the NP part can be studied, we do question it. The $P^2$ dependence can
be studied by means of a dispersion relation in the target ${\rm mass}^2 =
\mu^2$,  with threshold at $\mu^2 = 4 m_{\pi}^2$  \cite{Bj}, \cite{IW}.
The NP part of the structure function does have a contribution from small values
of the mass parameter in this dispersion relation. In particular, there
is a double pole at $\mu^2 = m_{\rho}^2$ corresponding to the simple $\rho$
dominance model which we have studied. However, the NP terms in the photon
structure function which are inverse powers of $P^2 \ll \ol^2$  are not in fact
so simple. We expect that the dominant terms behave as \cite{IW},
\begin{equation}
F_{2,NP}^\gamma (x,Q^2,P^2)
\simeq \frac{f_1(x,Q^2)}{(1 + \frac{P^2}{ m^{2}_{1} })^2}+
       \frac{f_2(x,Q^2)}{(1 + \frac{P^2}{ m^{2}_{2}})}.
\end{equation} 
In writing this, we have chosen to turn an expansion in powers of $1/P^2$  into
one which is well behaved as $P^2 \rightarrow 0$. This represents an assumption
about how higher order terms behave which we think is reasonable (but not proven).
Only one of these terms can be thought of as corresponding to the vector meson
double pole term, if $m_1 \rightarrow m_{\rho}$. The other might have physically
an interference term between a vector meson pole (for $m_2 
\rightarrow m_{\rho}$) and another term with no pole in $\mu^2$. We can see no
reason why the second term cannot be important. We do not know of theoretical
arguments which could establish its physical significance or size. 

We believe that these considerations make it clear why the $P^2$ dependence
of the photon structure function is worth careful experimental study. 
We hope that these experiments will be carried out. As yet, little has been done
in this promising area at the interface of perturbative and nonperturbative
quantum chromodynamics.

\section*{Acknowledgments}
This work was
supported in part by the U.S.~Department of Energy under Contract No.
DE-AC02-83ER40105. 

\newpage

%%%%%%%%%%%%%%%%%%%%%%%%%%%%%%%%%%%%%%%%%%%%%%%%%%%%%%%%%%%%%%%%%%%%%%%%%%%%
%..........................BIBLIOGRAPHY.................................
%%%%%%%%%%%%%%%%%%%%%%%%%%%%%%%%%%%%%%%%%%%%%%%%%%%%%%%%%%%%%%%%%%%%%%%%%%%%

\newpage

%%%%%%%%%%%%%%%%%%%%%%%%%%%%%%%%%%%%%%%%%%%%%%%%%%%%%%%%%%%%%%%%%%%%%%%%%%
%                          FIGURES
%%%%%%%%%%%%%%%%%%%%%%%%%%%%%%%%%%%%%%%%%%%%%%%%%%%%%%%%%%%%%%%%%%%%%%%%%%

\section*{Figures}
\begin{enumerate}

\item
\label{NPmoments}
This figure set shows moments of the NP part of the photon structure 
function at $P^2=0$, with $Q^2$ ranges from $Q_0^2=3$ to $45$ GeV$^2$
The moments for $n=2$, 3, 4 correspond to plots 1(a), 1(b), 1(c).
Each curve contains NP behavior of the photon structure function for 
the three fits A, B, C mentioned in the text.

\item
\label{realcase}
These plots show the total real photon structure functions for the three fits.
2(a) and 2(b) correspond to $n=4$, 6 moments respectively.
Each plot shows the corresponding total moment for the PQCD and NP structure
functions superimposed.
Dashed curves are PQCD parts and solid curves are total moments for the
examples A, B, C.

\item
\label{total}
The figures 3(a), 3(b), and 3(c) correspond to $n=2$, $n=4$, and $n=6$ moments
of the virtual photon structure function respectively. There are also $P^2$
dependence in the figures. $n=2$ case is the most interesting one since we can
not see this moment in the real photon structure function.
Dashed curves correspond to PQCD parts of the structure function
and solid curves to total moments for the examples, A, B, C.
 
\end{enumerate}


\begin{thebibliography} {0}
\bibitem{err} We found an extra flavor $f$ multiplied in the equation (A.17) in
  reference \cite{UW}.   Therefore
  $A^{(2)}_n=6(\la e^2\ra,0,\la e^4\ra -\la
  e^2\ra ^2) \tilde A^{(2) \psi}_{nG}$ should replace
  $A^{(2)}_n=6f(\la e^2\ra ,0,\la e^4\ra -\la e^2\ra ^2) \tilde A^{(2)
  \psi}_{nG}$.
\end{thebibliography}
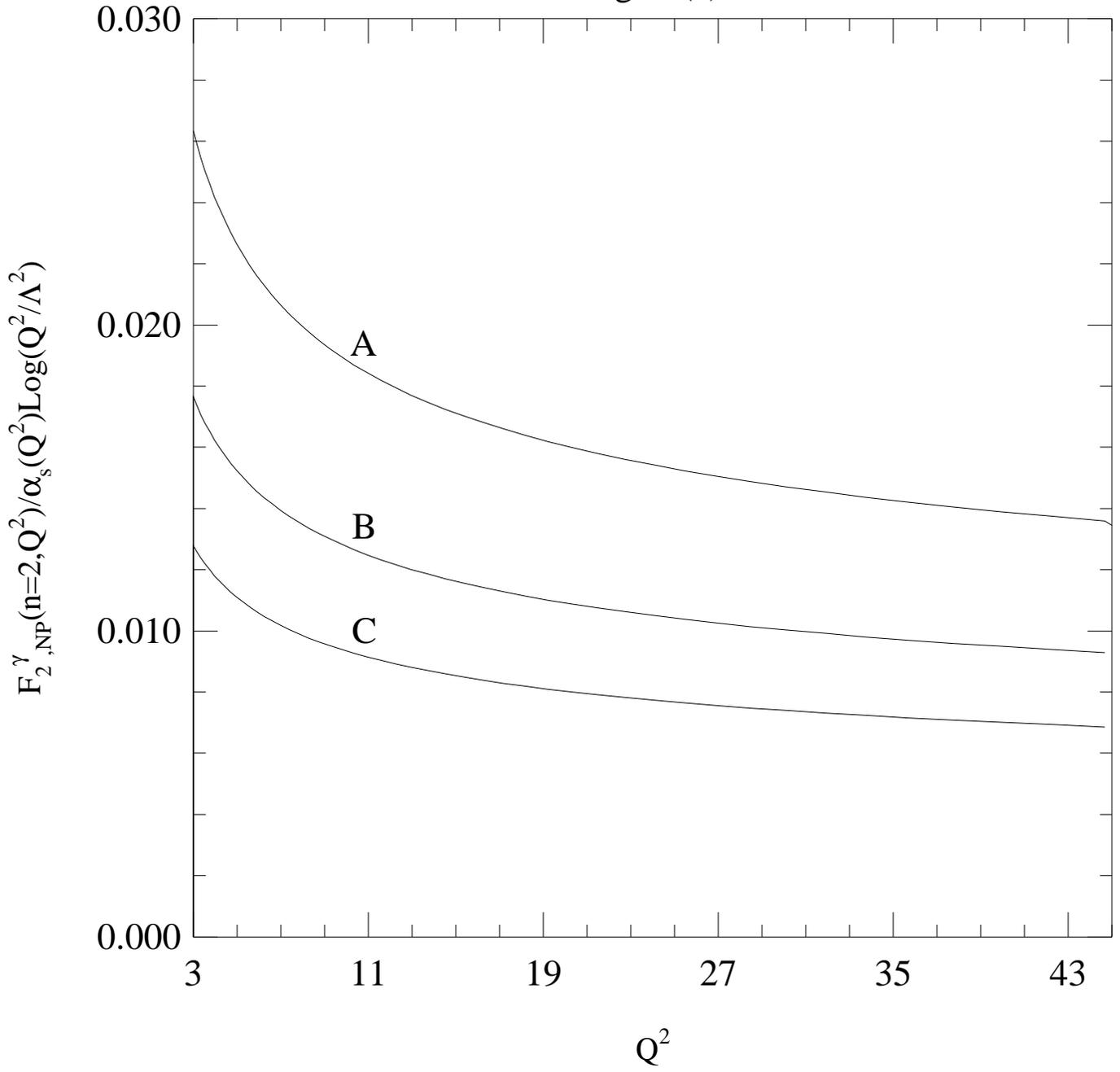
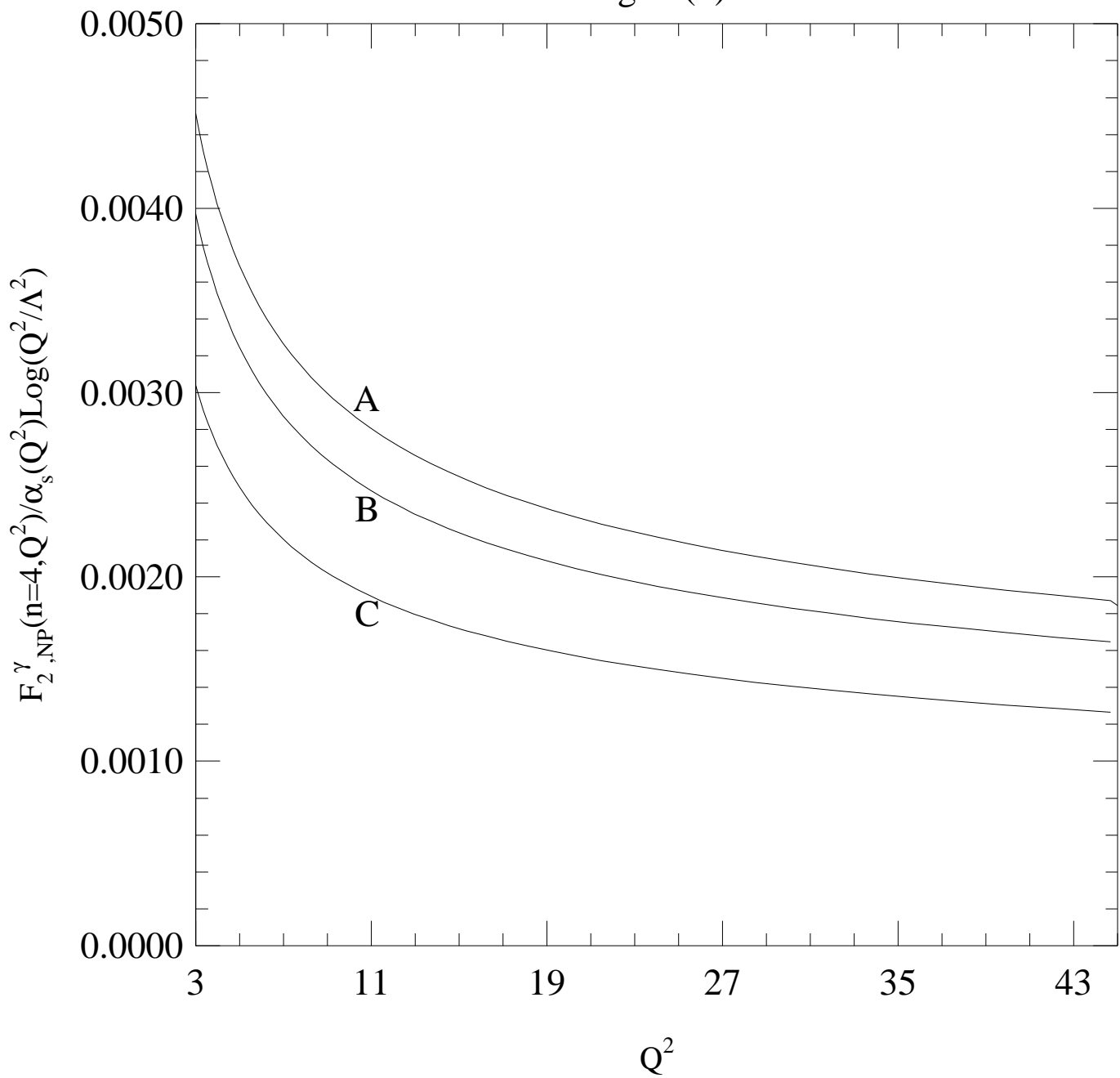
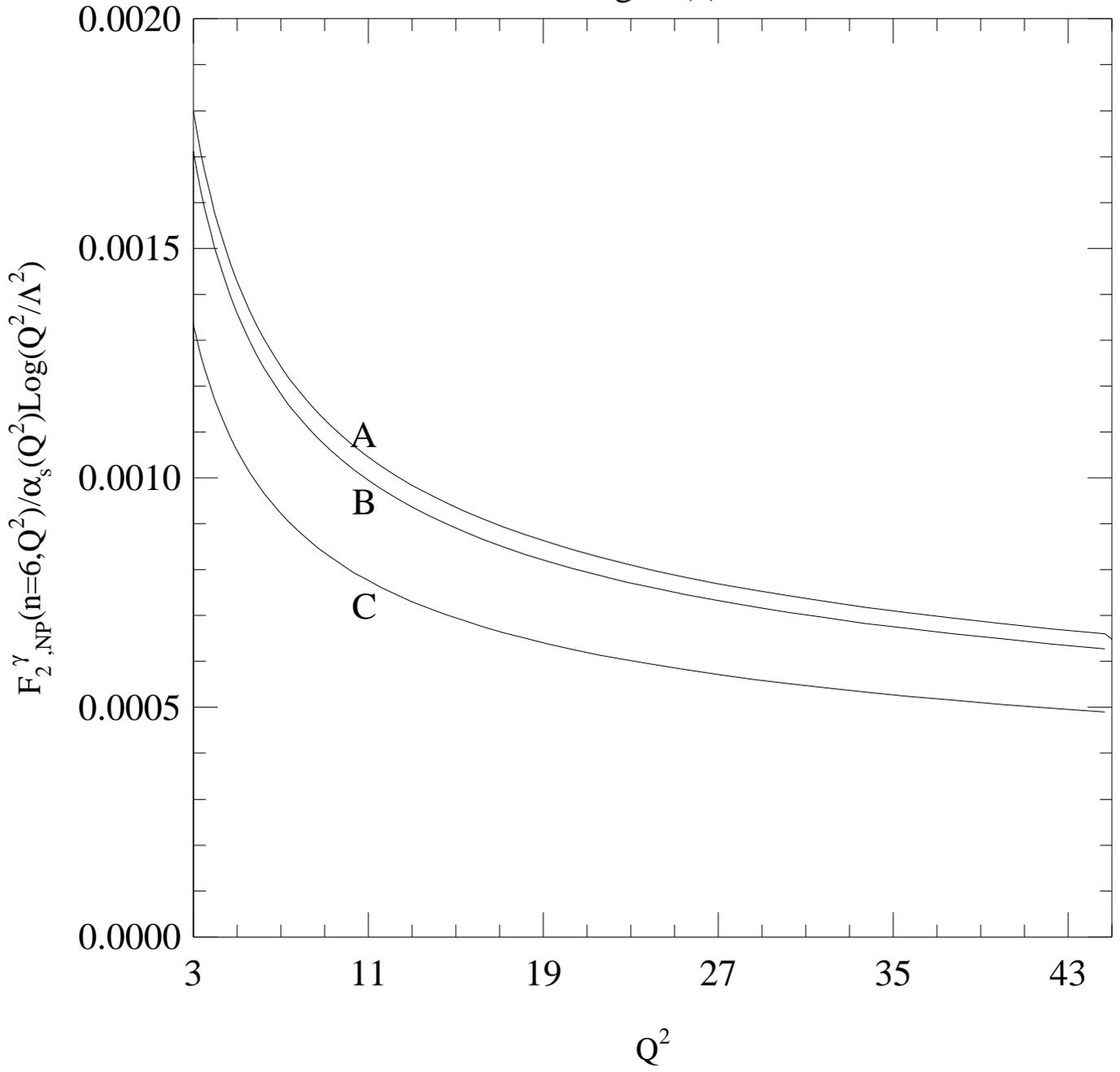
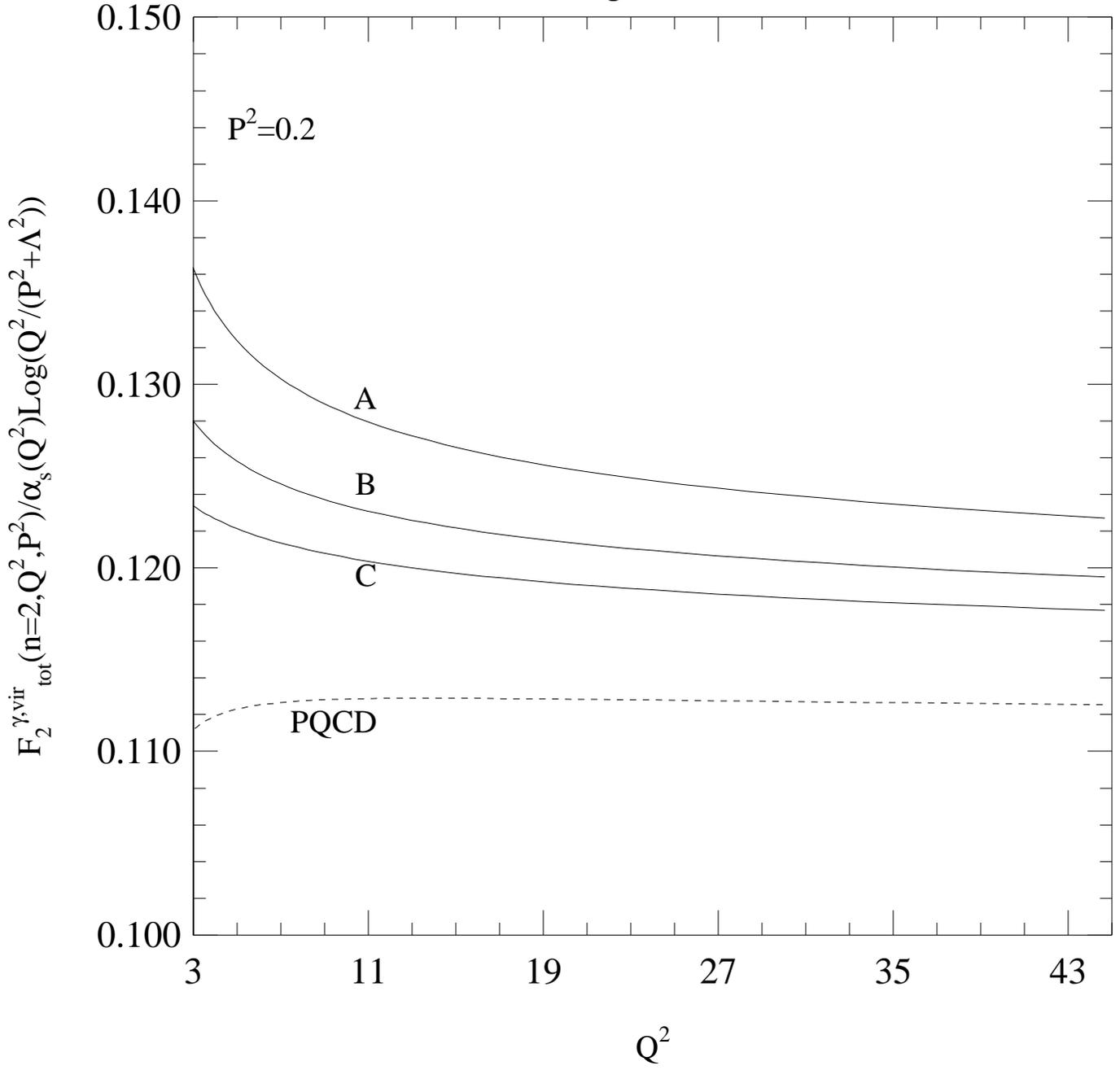
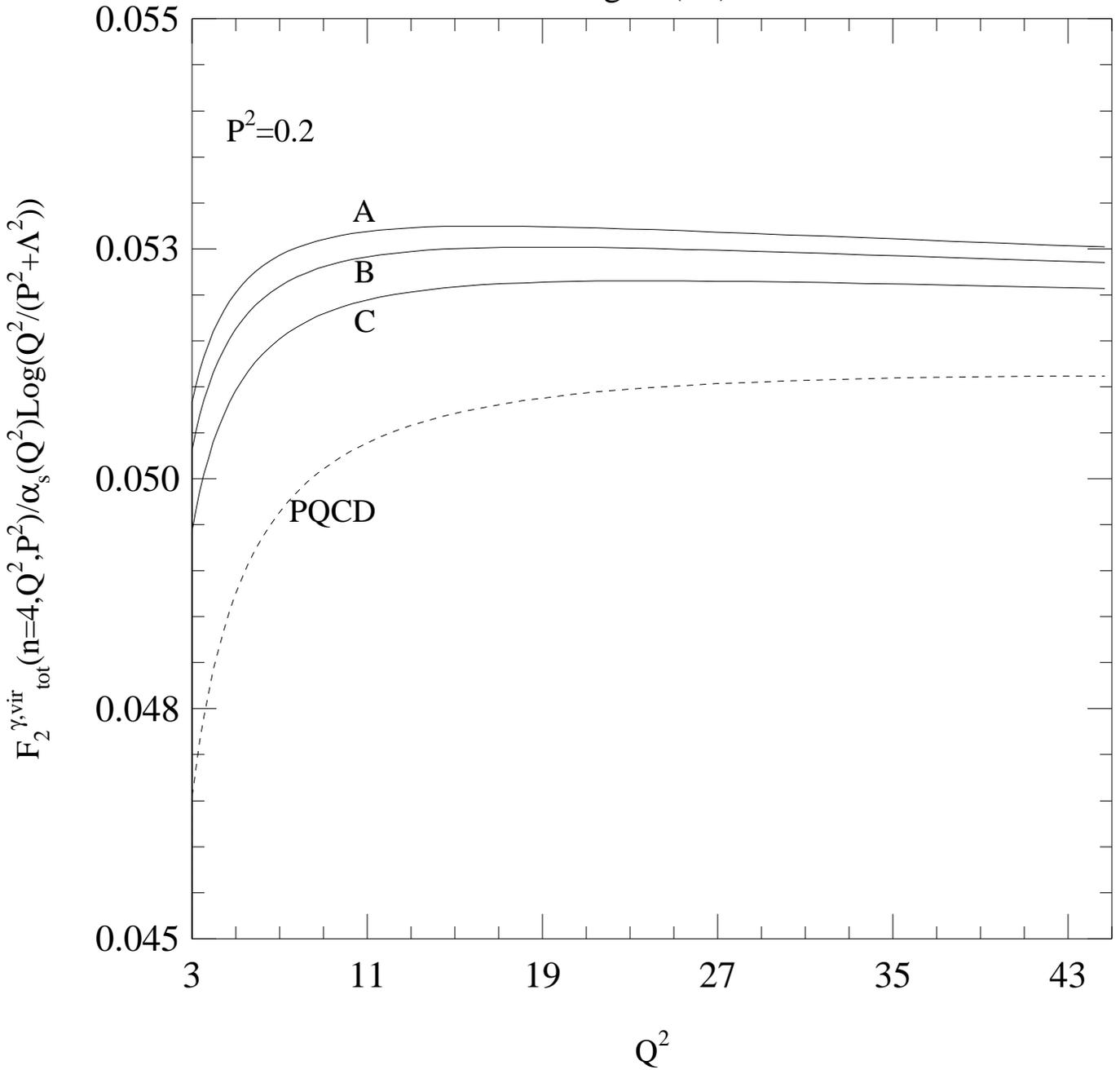
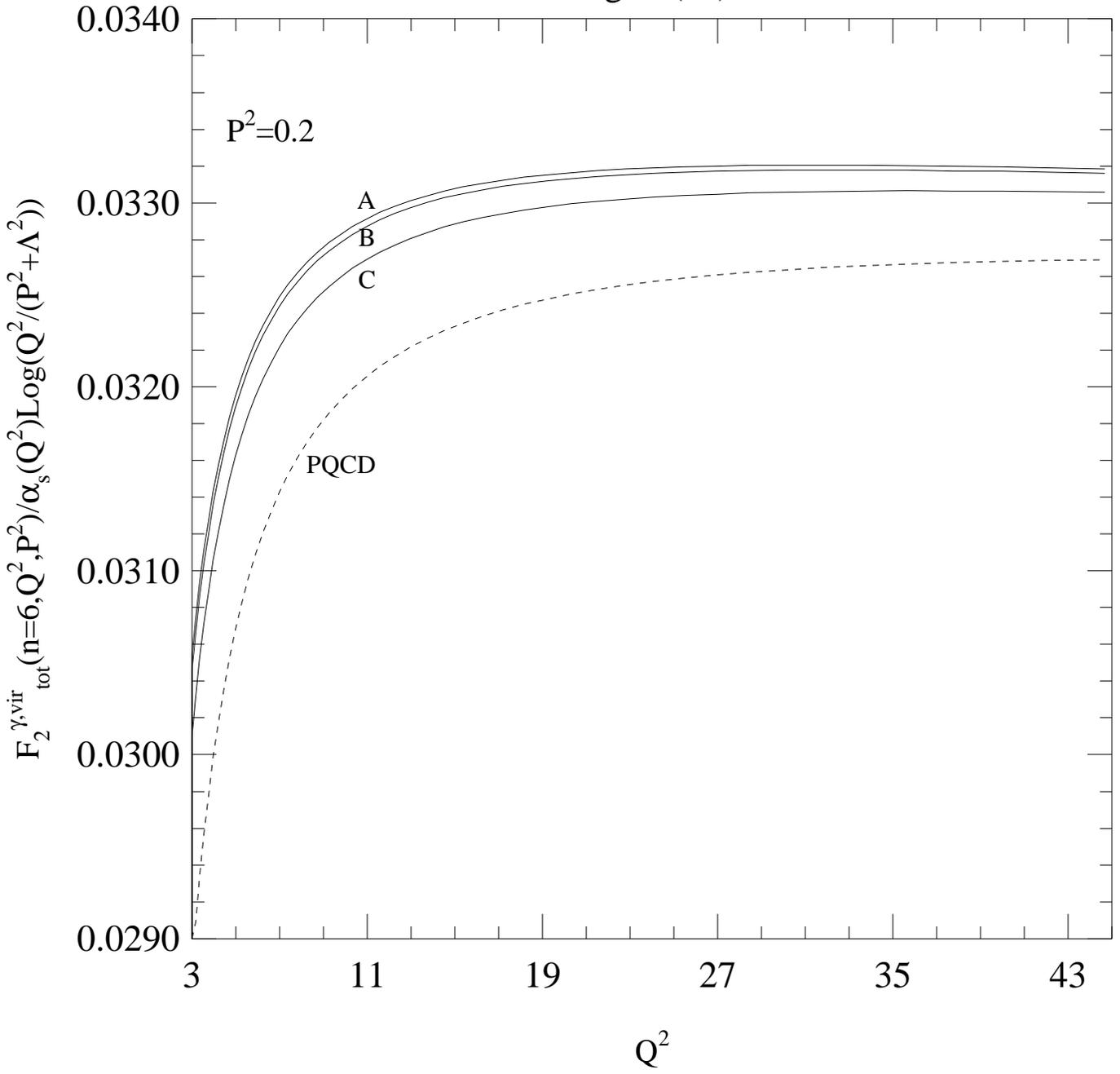
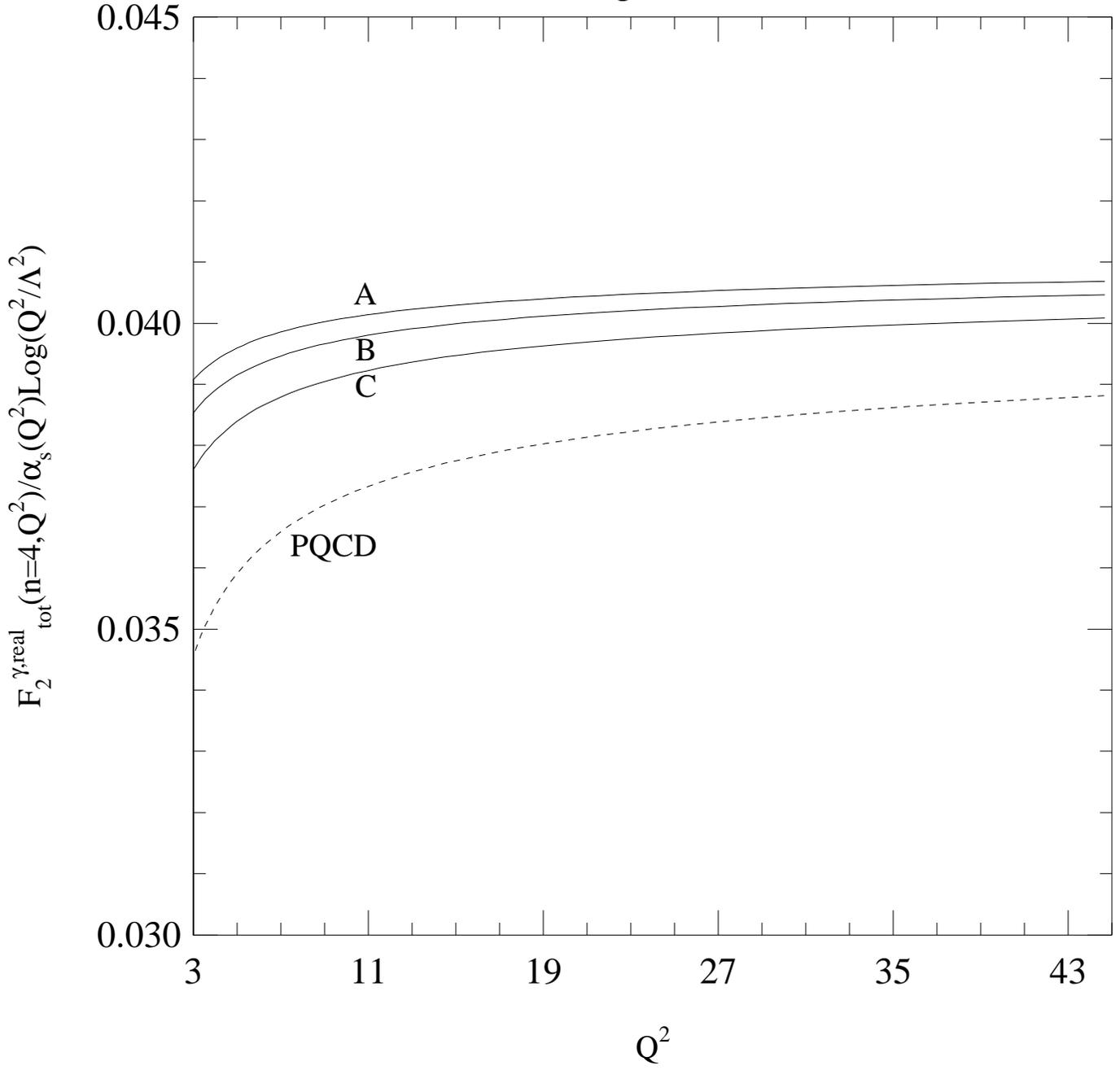
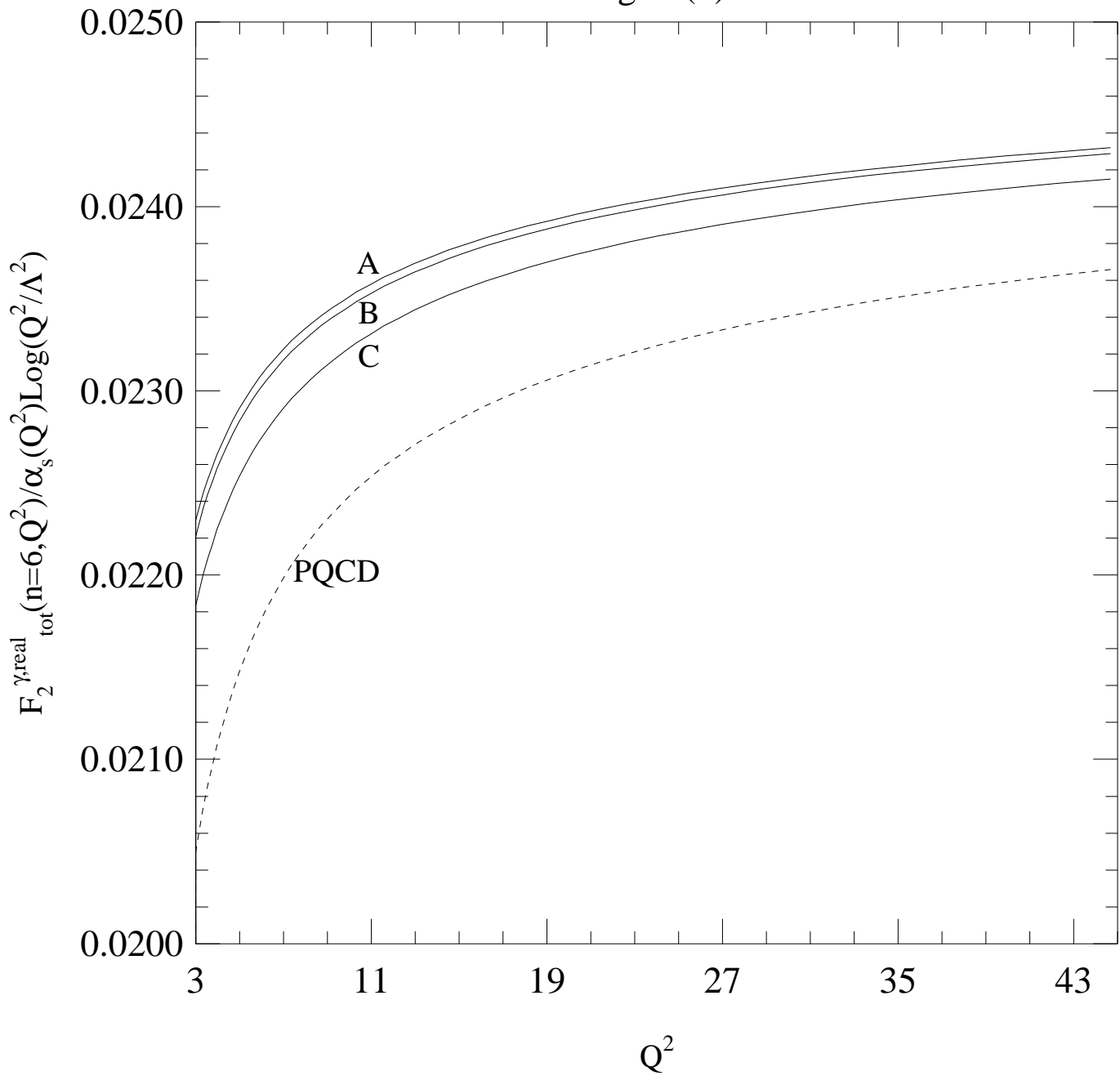
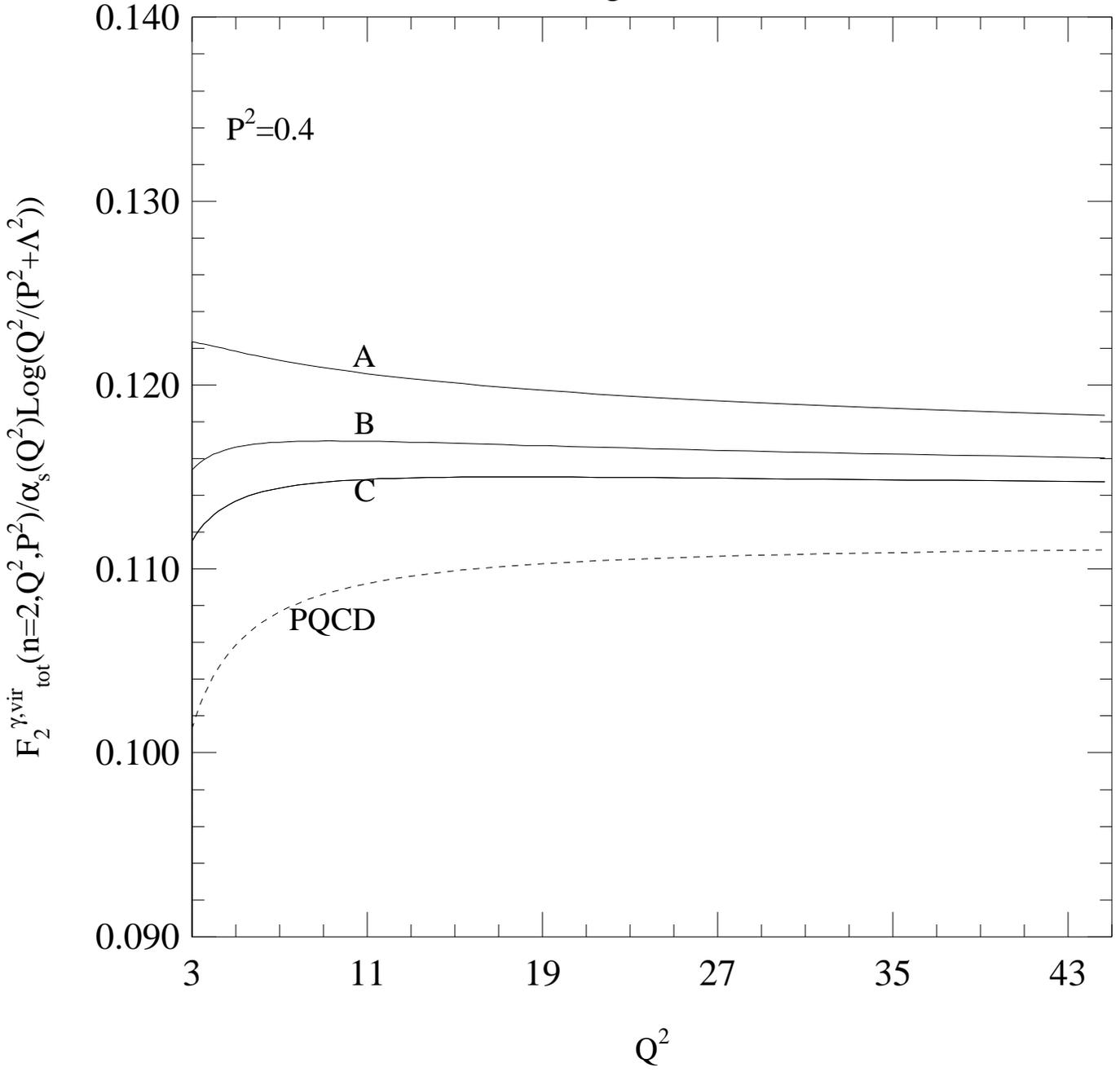
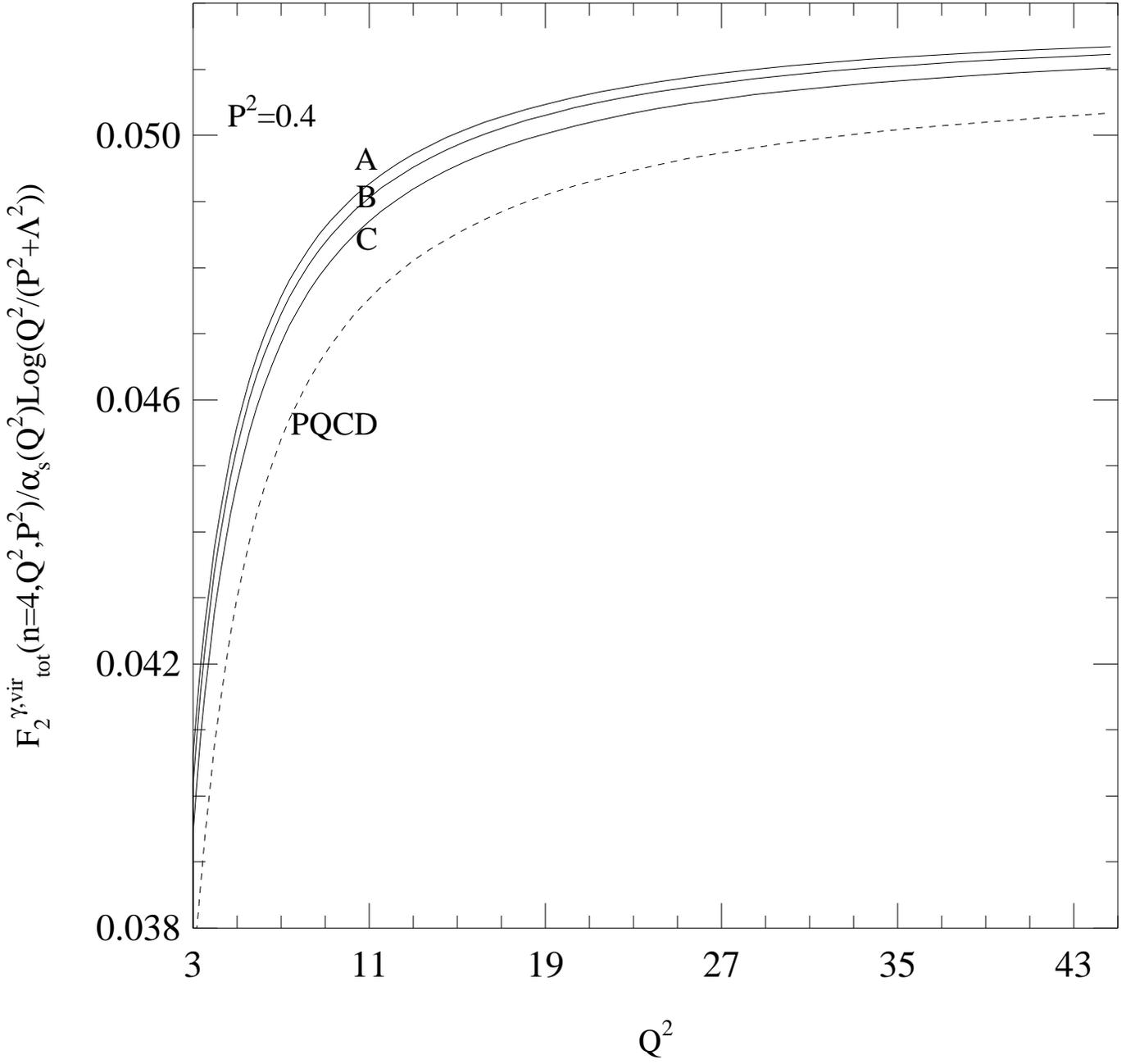
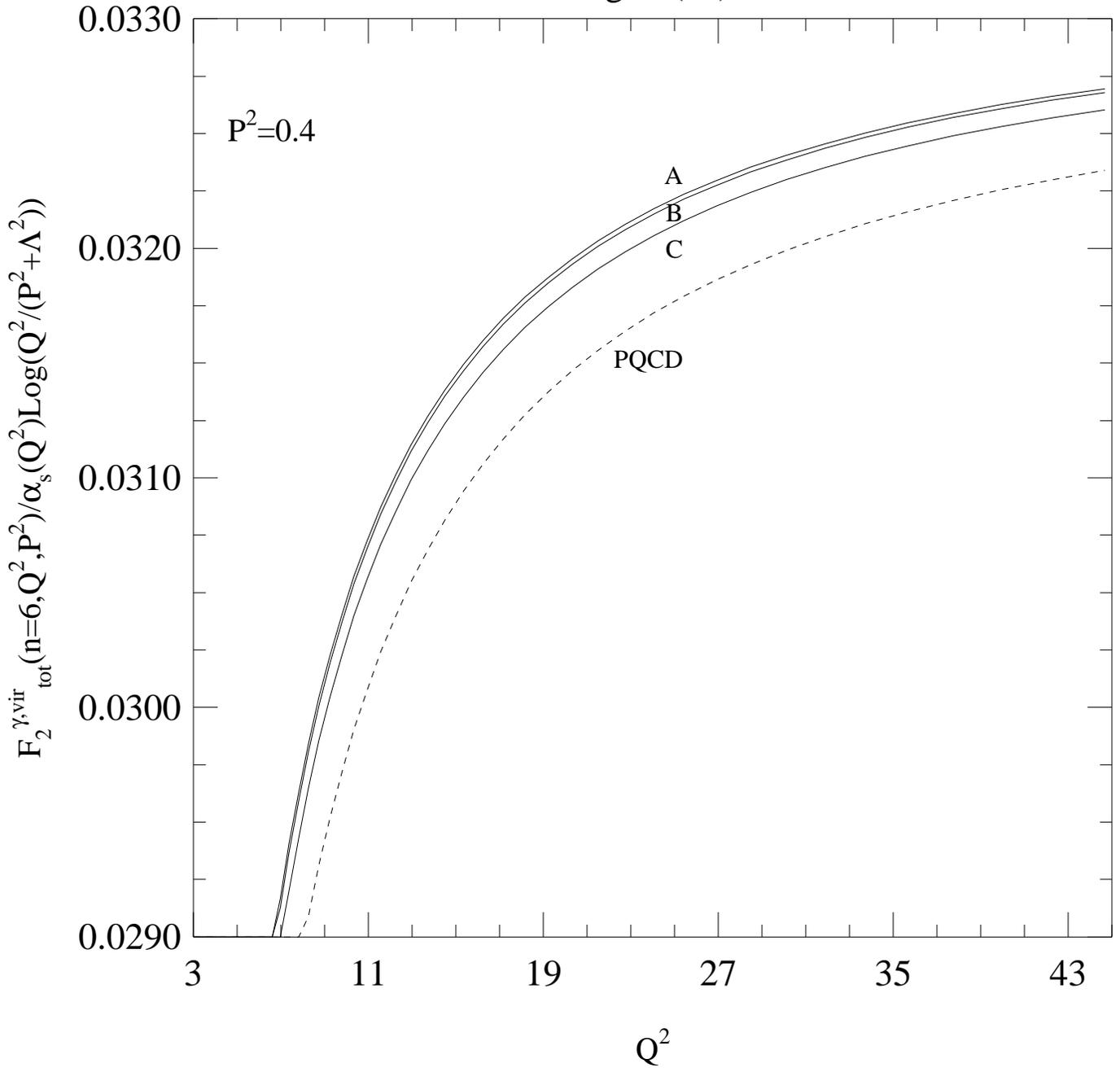
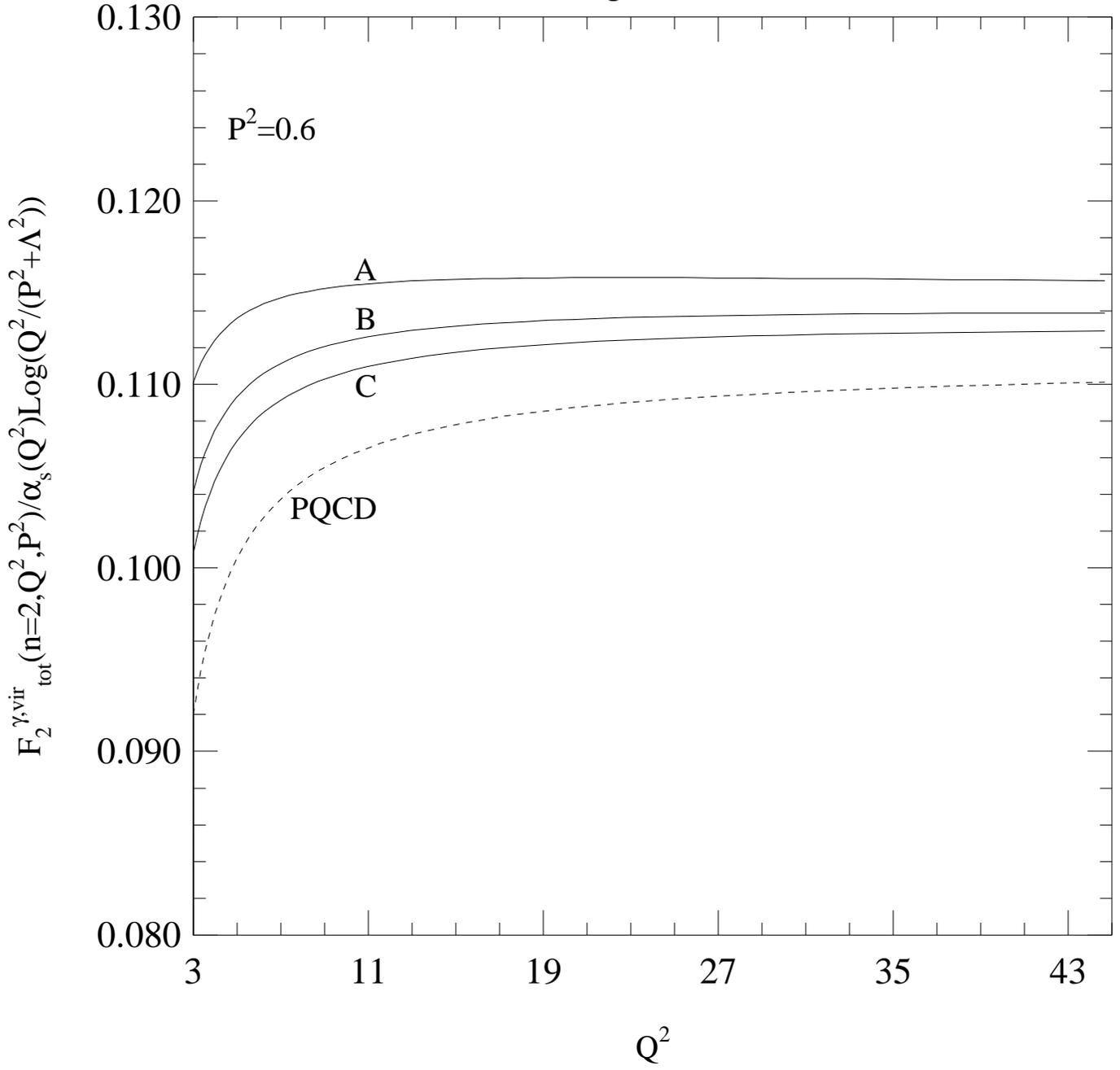
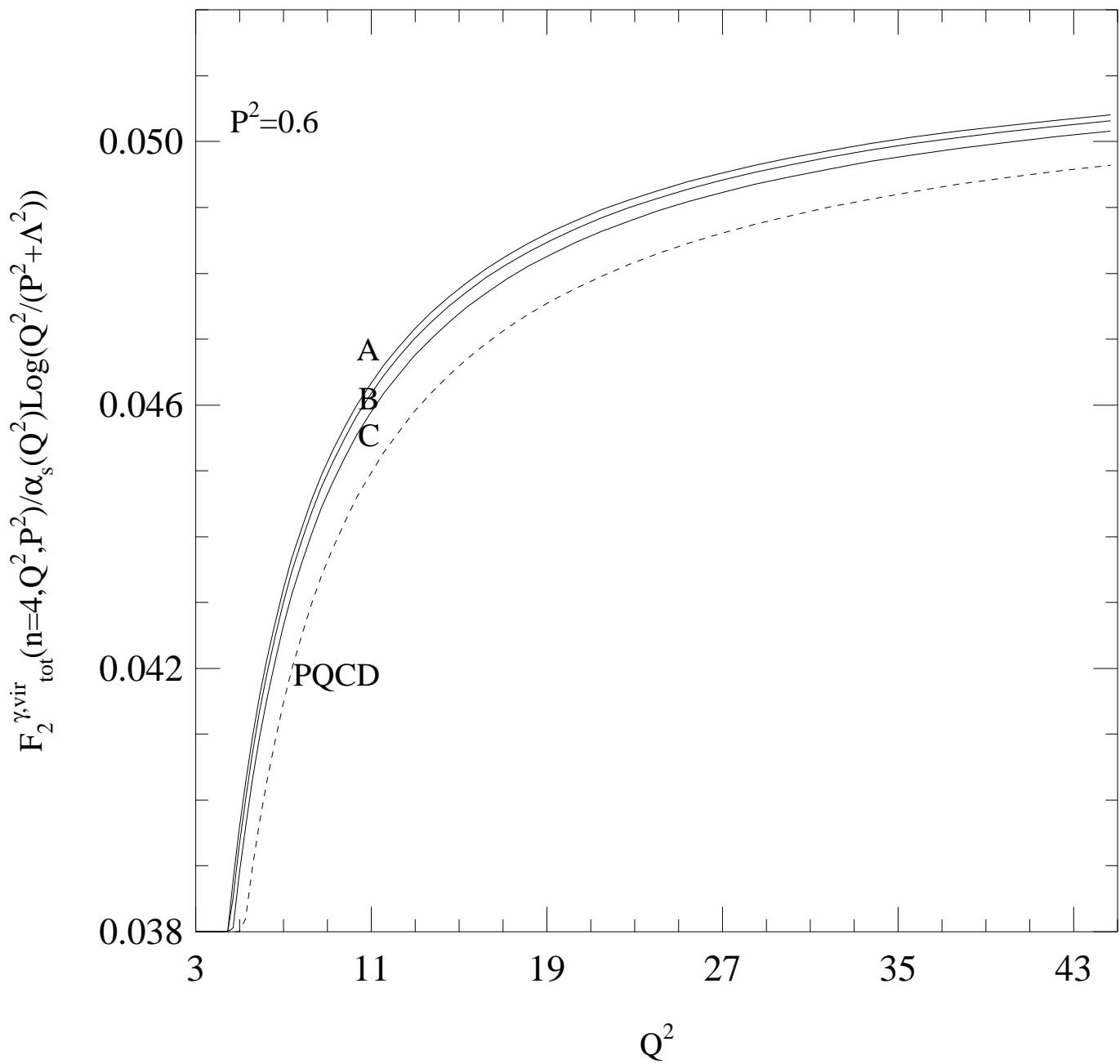
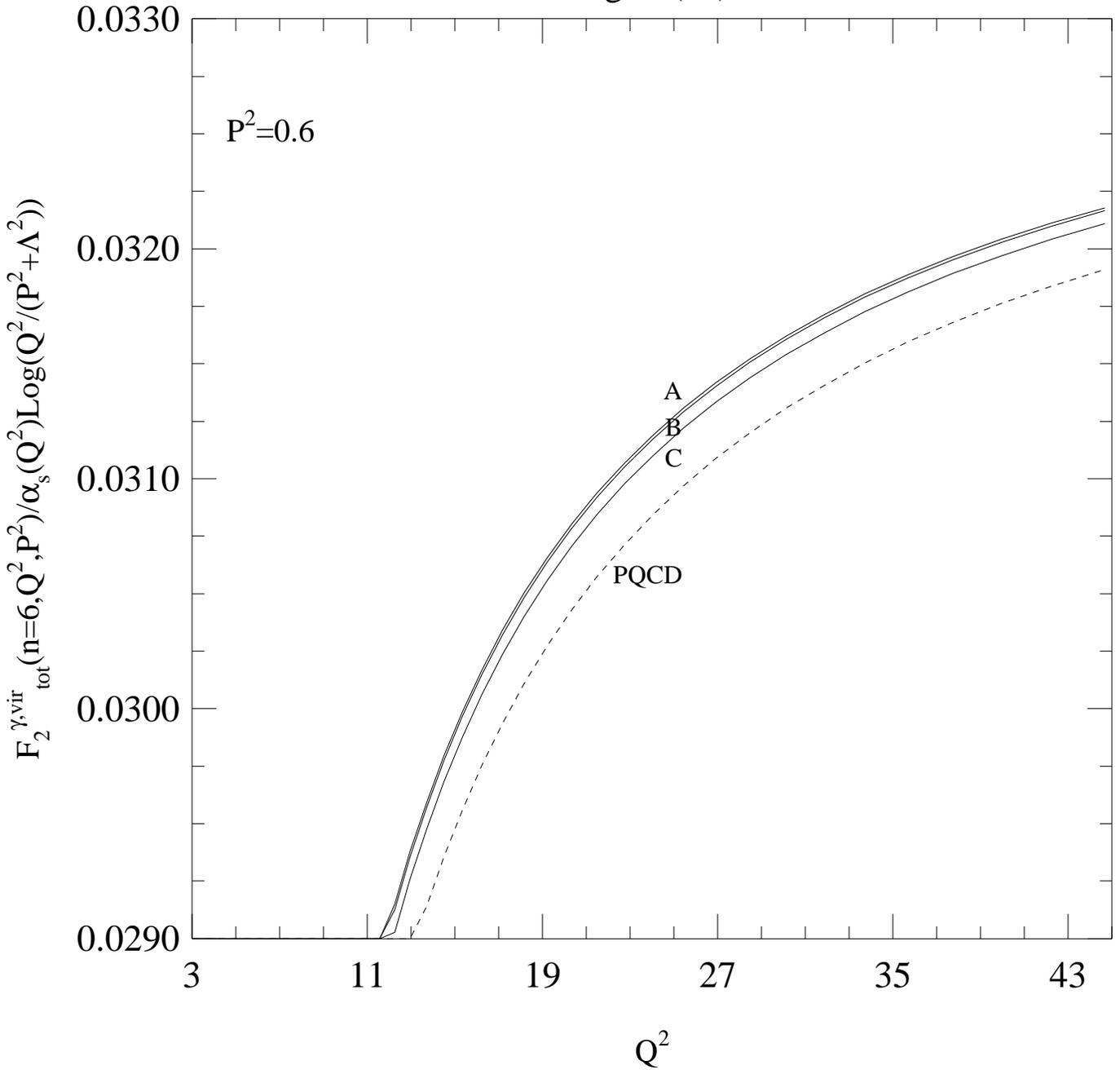
\end{document}